\documentclass[conference]{IEEEtran}
\IEEEoverridecommandlockouts
\usepackage{graphicx, amsmath, amsfonts,multirow, amssymb, tikz, algorithm, algorithmic, caption, lipsum, cuted, soul, graphics}%, algpseudocode}
\usepackage{mathtools}
\usepackage{pifont}% http://ctan.org/pkg/pifont malik this package & following lines to use check marks
\usepackage{array}
\usepackage[english]{babel}
\usepackage{epstopdf}
\usepackage{cite}
\usepackage{longtable}
\usepackage{pdflscape}
\usepackage{graphicx}
\usepackage{url}
\usepackage[pscoord]{eso-pic} % malik IFIP Copyright Clearance Code Notice
\usepackage{hyperref}%[hidelinks]{hyperref}  % malik [hidelinks]/[draft] hide/disable links for everything and generates bookmarks
\hypersetup{ colorlinks=true, citecolor=blue,  linkbordercolor=blue}
\usepackage[depth=subsection]{bookmark} % malik this creates links for everything and generates bookmarks
\usepackage{float}
\usepackage[export]{adjustbox}
\usepackage{color}% This package to use colored writting
\usepackage{subcaption}% for multi part pictures
\usepackage{multirow} % Malik this line for multi rows tables
\usepackage{textcomp} % malik to show astrophe
\usepackage{titlesec}
\usepackage{float} % malik to force figures/algorithms/tables to a particular place
\usepackage[switch]{lineno} % malik these two lies are used for numbering pdf lines
\usepackage[figurename=Fig.]{caption} % malik these two lines to make Fig. instead of Figure

\titleformat{\paragraph}
{\normalfont\normalsize\bfseries}{\theparagraph}{1em}{}
\titlespacing*{\paragraph}
{0pt}{3.25ex plus 1ex minus .2ex}{1.5ex plus .2ex}

\def\be{\begin{equation}}
\def\ee{\end{equation}}
\def\ba{\left[\begin{array}}
\def\ea{\end{array}\right]}

\usepackage{inputenc}
\usepackage{pgfplots}
\usepgfplotslibrary{groupplots,dateplot}
\usetikzlibrary{patterns,shapes.arrows}
\DeclareRobustCommand{\IEEEauthorrefmark}[1]{\smash{\textsuperscript{\footnotesize #1}}}

\begin{document}

\title{Hierarchical Multi-Agent DRL-Based Framework for Joint Multi-RAT Assignment and Dynamic Resource Allocation in Next-Generation HetNets}

\author{\IEEEauthorblockN{Abdulmalik Alwarafy\IEEEauthorrefmark{1}, Bekir Sait Ciftler\IEEEauthorrefmark{1}, \IEEEmembership{Member, IEEE,} Mohamed Abdallah\IEEEauthorrefmark{1}, \IEEEmembership{Senior Member, IEEE,} \\
Mounir Hamdi\IEEEauthorrefmark{1}, \IEEEmembership{Fellow Member, IEEE,} and Naofal Al-Dhahir\IEEEauthorrefmark{2}, \IEEEmembership{Fellow Member, IEEE}}  \\
\vspace{-.4cm}
\IEEEauthorblockA{\IEEEauthorrefmark{1}Division of Information and Computing Technology, College of Science and Engineering, \\ Hamad Bin Khalifa University, Doha, Qatar} \\
\vspace{-.4cm}
\IEEEauthorblockA{\IEEEauthorrefmark{2}Electrical and Computer Engineering Department,
Erik Jonsson School of Engineering and Computer Science, \\The University of Texas at Dallas, USA}
%Emails: \{aalwarafy, bciftler, moabdallah, mhamdi\} {@hbku.edu.qa}
\vspace{-.64cm}
\thanks{A conference version of this work was published in the IEEE International
Conference on Communications Workshops (ICC Workshops) 2021 Proceedings \cite{alwarafy2021DeepRAT}.}
%\thanks{ - Abdulmalik Alwarafy is with the Division of Information and Computing Technology, College of Science and Engineering, Hamad Bin Khalifa University, Doha, Qatar. E-mail: aalwarafy@hbku.edu.qa}
}

\maketitle

\begin{abstract}
This paper considers the problem of cost-aware downlink sum-rate maximization via joint optimal radio access technologies (RATs) assignment and power allocation in next-generation heterogeneous wireless networks (HetNets). We consider a future HetNet comprised of multi-RATs and serving multi-connectivity edge devices (EDs), and we formulate the problem as a mixed-integer non-linear programming (MINP) problem. Due to the high complexity and combinatorial nature of this problem and the difficulty to solve it using conventional methods, we propose a hierarchical multi-agent deep reinforcement learning (DRL)-based framework, called DeepRAT, to solve it efficiently and learn system dynamics. In particular, the DeepRAT framework decomposes the problem into two main stages; the \textit{RATs-EDs assignment stage}, which implements a single-agent Deep $Q$ Network (DQN) algorithm, and the \textit{power allocation stage}, which utilizes a multi-agent Deep Deterministic Policy Gradient (DDPG) algorithm. Using simulations, we demonstrate how the various DRL agents efficiently interact to learn system dynamics and derive the global optimal policy. Furthermore, our simulation results show that the proposed DeepRAT algorithm outperforms existing state-of-the-art heuristic approaches in terms of network utility. Finally, we quantitatively show the ability of the DeepRAT model to quickly and dynamically adapt to abrupt changes in network dynamics, such as EDs' mobility.  
\end{abstract}


\vspace{-.24cm}
\section{Introduction} \label{Introduction}
Heterogeneous wireless networks (HetNets) are expected to be one of the key enablers for next-generation wireless communication networks \cite{RN234}. In such networks, a massive number of multi-radio access technologies (multi-RATs) in the licensed and unlicensed frequency bands across the ground, space, and underwater coexist to enhance the network's quality of service (QoS). The main goal of next-generation HetNets is to support the stringent QoS requirements of the emerging disruptive wireless applications in terms of rate, coverage, and reliability with ubiquitous connectivity. On the other hand, emerging user edge devices (EDs) are equipped with advanced multi-access physical capabilities that enable them to simultaneously aggregate radio resources from various RATs (i.e., multi-homing mode of operation) to guarantee an enhanced reliable communication for their running applications \cite{ciftler2021dqn}. The number of such EDs is expected to be around 30 Billion by 2023, including smartphones, IoT devices, and sensors \cite{Cisco_report}.

Radio resource allocation is crucial in the planning, orchestration, and resource optimization of next-generation HetNets. It is mainly used to guarantee enhanced system efficiency, increased network connectivity, and reduced energy consumption. However, allocating and managing radio resources, such as power, spectrum, and rate, in the next-generation HetNets is a persistent challenge. In particular, the multi-RAT assignment for EDs (i.e., RATs-EDs associations) and RATs' power allocation are among the key issues in the emerging HeNets. Various methods have been proposed to address these two radio resource allocation issues, such as optimization theory, ranking-based, and game theory \cite{hussain2020machine, alwarafy2021deep, chkirbene2021deep}. However, most of these conventional resource allocation methods generally suffer from the following shortcomings. They require full and real-time knowledge of the network dynamics. Unfortunately, obtaining such information is not possible as most real-world networks are dynamic and change over time causing a rapid variation of the wireless channel. In addition, most of these conventional resource allocation methods suffer from high computational complexity, lack of scalability, and do not always guarantee convergence. These issues are even exacerbated in next-generation HetNets due to the large-scale and massive heterogeneous nature of these networks in terms of the underlying RATs and QoS demands of supported applications as well as the explosive number and type of emerging EDs. All of the above mentioned issues and challenges render the use of existing resource allocation techniques for future wireless HetNets quite difficult if not even impossible \cite{RN1}. Hence, it is of a paramount importance to develop alternative resource allocation solutions that can overcome these challenges while quickly adapting to the varying systems’ dynamics.

Deep reinforcement learning (DRL) has emerged recently as one of the most promising branches of the artificial intelligence (AI) field. In DRL, intelligent agents are trained to make autonomous decisions and observe their results in order to learn optimal control policies. In the context of radio resource allocation, DRL methods possess several advantages over state-of-the-art techniques. First, they provide autonomous and real-time decision-making even for highly complex and large-scale HetNets. Second, DRL provides efficient solutions for complex and high-dimensional wireless radio resource allocation optimization problems with limited channel state information (CSI) knowledge. These unique features make DRL techniques one of the key enabling technologies that can be utilized to address the radio resource allocation in next-generation HetNets.  

In light of the imperative need for radio resource allocation in next-generation HetNets, the shortcomings of traditional radio resource allocation approaches, and the efficiency of DRL techniques in solving complex radio resource allocation optimization problems, this paper proposes a DRL-based framework for radio resource allocation in next-generation HetNets. In particular, we propose a hierarchical DQN and DDPG-based scheme called DeepRAT to study the problem of adaptive multi-RAT assignment and continuous downlink power allocation for multi-homing EDs in next-generation HetNets. Our aim is to jointly optimize the sum rate of the entire network, monetary cost, and power allocation while satisfying the QoS demands of EDs and the constraints of the RATs' power resources. Our simulation results demonstrate that the DeepRAT algorithm can efficiently learn the optimal policy, and it outperforms the greedy, random, and fixed algorithms in terms of satisfying the objective of the optimization problem. Moreover, after training, our proposed algorithm converges 2.5 times faster than the case with the initial training. In general, the main contributions of this paper are summarized as follows:

\begin{itemize}
    \item We formulate an optimization problem whose objective is to cost-effectively maximize the downlink sum-rate of the multi-RAT HetNet via jointly optimizing the RATs-EDs assignment and RATs' power allocations while considering the limited RATs' power resources, multi-homing capabilities of EDs, and QoS data requirements of EDs.
    \item Due to the extensive computational complexity and combinatorial nature of the formulated problem, as well as the difficulty of applying conventional approaches to solve it, we propose a DRL-based algorithm called DeepRAT to solve the problem hierarchically and learn the system dynamics using a mix of value-based and policy-based DRL algorithms. 
    \item Using simulations, we show how the various agents of the DeepRAT model interact in order to learn the global optimal policy and solve our proposed optimization problem, relying only on limited information about the network dynamics and CSI.
    \item We demonstrate quantitatively that our proposed algorithm outperforms existing heuristic-based methods in terms of satisfying the objective of the optimization problem. Also, we show how the DeepRAT algorithm can quickly adapt to the abrupt changes in network dynamics, such as EDs' mobility.
\end{itemize}

The rest of this paper is organized as follows. Table \ref{Acronyms} defines the main acronyms used in this paper. Section \ref{Related_Work} describes some related work that implements DRL methods for radio resource allocation. Section \ref{System_Model_Problem_Formulation} presents the proposed system model and formulates the optimization problem. Section \ref{DeepRAT_Architcutre} shows the architecture of the proposed DeepRAT framework and explains its underlying DRL-based models along with a brief mathematical background. Section \ref{sec4} explains the simulation setup and discusses the corresponding numerical results. Finally, Section \ref{sec6} concludes the paper.

\begin{table*}[htbp]
\caption{\textsc{Definitions of Main Acronyms Used in this Paper.}}
\centering
\label{Acronyms}
\begin{tabular}{|c|c||c|c||c|c|}
\hline
%\rowcolor[HTML]{} 
\textbf{Acronym} & \textbf{Definition} & \textbf{Acronym} & \textbf{Definition}& \textbf{Acronym} & \textbf{Definition} \\ \hline
RAT & Radio Access Technology& HetNet & Heterogeneous Network& ED& Edge Device\\ \hline
MINLP & Mixed-Integer Non-Linear Programming & DRL& Deep Reinforcement Learning& DQN & Deep Q Network \\ \hline
DDPG& Deep Deterministic Policy Gradient& QoS& Quality of Service & AI& Artificial Intelligence \\ \hline
OFDMA& Orthogonal Frequency Division Multiple Access& CSI & Channel State Information & V2V& Vehicle-to-Vehicle \\ \hline
NEC& Neural Episodic Control& VLC& Visible Light Communication & D2D& Device-to-Device \\ \hline
DSRC & Dedicated Short-Range Communication& PEN& Patient Edge Node& RF& Radio Frequency \\ \hline
SDN & Software-Defined Networking& DNN& Deep Neural Network& ES& Edge Server \\ \hline
AWGN& Additive White Gaussian Noise& SNR& Signal to Noise Ratio& AP& Access Point  \\ \hline
C-RAN& Cloud/Centralized Radio Access Network& LTE&  Long-Term Evolution& BBU & Baseband Unit \\ \hline
CDF& Cumulative Distribution Function & UAV& Unmanned Aerial Vehicle& OU& Ornstein-Uhlenbeck \\ \hline
OFDM& Orthogonal Frequency Division Multiplexing  & IoT& Internet of Things & NR& New Radio \\ \hline
\end{tabular}
\vspace{-4mm}%Put here to reduce too much white space after your table
\end{table*}

\section{Related Work} \label{Related_Work}
DRL-based techniques have attracted considerable research lately in the context of radio resource allocation for wireless networks \cite{alwarafy2021deep,RN1}. The authors in \cite{khan2020centralized} proposed DRL models based on the single and multi-agent actor-critic algorithms to address the problem of total sum-rate maximization via power allocation for cellular networks. In \cite{RN297}, the authors used DRL methods to study the joint optimization of user association and power allocation in orthogonal frequency division multiple access (OFDMA)-based HetNets. Ye \textit{et al.} \cite{RN14} presented a DRL-based mechanism to study the problem of resource allocation for unicast and broadcast scenarios in vehicle-to-vehicle (V2V) networks. The work in \cite{bi2020deep} investigated the power control problem of device-to-device (D2D)-enabled networks in time-varying environments using a centralized DRL algorithm. Zhang \textit{et al.} \cite{zhang2020deep} presented a DRL algorithm to study the problem of energy-efficient resource allocation in ultra-dense cellular networks. The authors in \cite{munaye2021deep} presented a multi-agent DQN-based model to study the problem of joint power, bandwidth, and throughput allocation in unmanned aerial vehicle (UAV)-assisted IoT networks. In \cite{ciftler2021dqn}, the authors proposed a non-cooperative multi-agent DQN-based method to study the problem of power allocation in hybrid RF/VLC networks. The authors show via simulation that the convergence rate of the DQN-based model is 96.1\% compared to that of the $Q$-learning-based algorithm, which is 72.3\%.   

On the other hand, using deep deterministic policy gradient (DDPG) models has also gained increasing interest recently. They have shown superior performance in addressing the radio resource allocation problems in continuous and high dimensionality environments compared to the vanilla DQN algorithms\cite{sutton2018reinforcement}. The authors in \cite{RN295} presented a comparative study for the applications of three DRL algorithms, namely the DDPG, Neural Episodic Control (NEC), and Variance Based Control, in the optimization of wireless networks. The authors concluded that the DDPG and VBC methods achieve better performance than the NEC-based algorithm. In \cite{chkirbene2021deep}, the authors presented a single-agent DDPG algorithm to address the problem of network selection in heterogeneous health systems. Their goal was to optimize the medical data delivery from Patient Edge Nodes (PENs) via multi-radio access networks (RANs) to the core network. Nasir \textit{et al.} \cite{nasir2020deepjoint} presented a multi-agent DDPG-based algorithm to study the problem of joint power and spectrum allocation in wireless networks. Based on simulation results, the authors demonstrated how their proposed technique outperforms the conventional fractional programming algorithm. In \cite{RN296}, the authors investigated the problem of rate resource allocation for 5G network slices. The authors decomposed the problem into a master-slave, and proposed a multi-agent DDPG-based algorithm to solve it. Experimental results showed that their proposed algorithm performs better than some baseline approaches and provides a near-optimal solution.

In this paper, we present a multi-agent algorithm based on DRL called the DeepRAT to study the problem of cost-effective sum-rate maximization of HetNets via dynamic multi-RAT assignment and continuous power allocation for multi-connectivity multi-homing EDs. Towards this end, we formulate this problem as a mixed-integer non-linear programming (MINLP) problem and, due to the high complexity and combinatorial nature of the problem, we propose the DeepRAT algorithm to solve it efficiently and learn system dynamics, relying only on limited information about the network and CSI.

%Malik Fig
\begin{figure*}[hbt!]
	\centering
	%\vspace{-.5cm}
	\includegraphics[width=17.4cm, height=10cm,center]{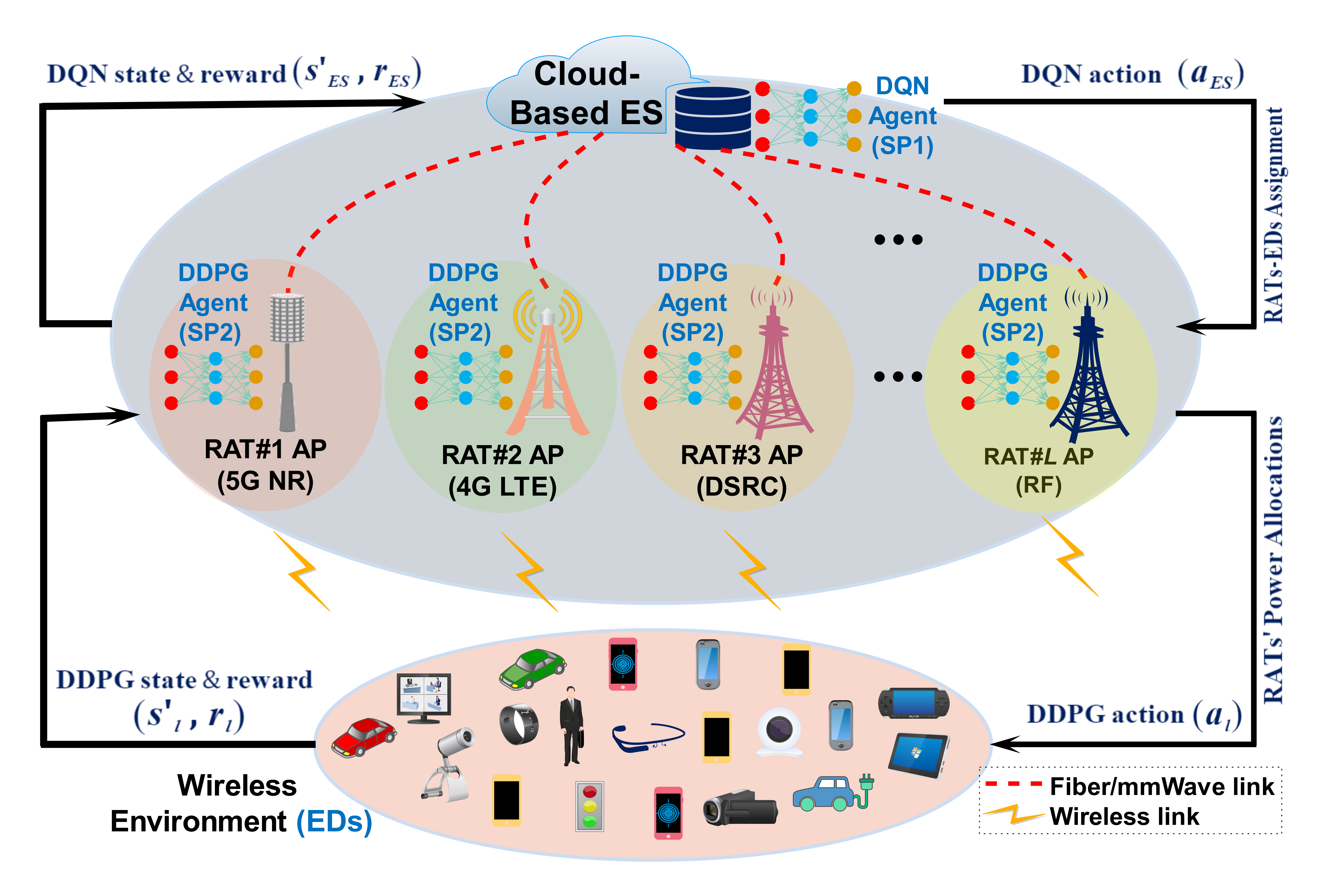}
	\caption{System model of our proposed DeepRAT framework for next-generation HetNets. It comprises $L$ radio access technology (RAT) access points (APs) controlled by a cloud-based edge server (ES) and serves $U$ multi-homing edge devices (EDs).}
	\label{Fig1}
\vspace{-4mm}%Put here to reduce too much white space after your table
\end{figure*}

%\squeezeup
%\vspace{-0.35cm}
\section{System Model and Problem Formulation} \label{System_Model_Problem_Formulation}
This section describes our proposed system model and formulates the optimization problem. 
%\squeezeup
%\vspace{-.2cm}

\subsection{System Model} \label{subsec2-1}
We consider a next-generation HetNet as depicted in Fig. \ref{Fig1}. It consists of various RAT access points (APs), such as sub 6GHz, dedicated short-range communication (DSRC) for vehicular networks, 5G NR, 4G long-term evolution (4G LTE), and WiFi. It is assumed that the RATs have different operating characteristics, such as carrier frequency, spectrum, data rate, energy consumption, the monetary cost for using RAT services, and transmission delay. In order to guarantee judicious and efficient management of networks radio resources, we assume that the RATs are controlled by a cloud-based edge server (ES). The RATs are assumed to serve multi-connectivity (i.e., multi-access), multi-homing EDs. Note that, unlike our previous work in \cite{alwarafy2021DeepRAT} in which each ED can be assigned one RAT at any time in a greedy fashion, i.e., multi-mode, the EDs in this paper are assumed to have the ability to connect to multiple RATs at any time to aggregate RATs' radio resources.
%\vspace{-.2cm}

\subsection{Optimization Problem Formulation} \label{subsec2-2}

\begin{table}[bt!]
\centering
\caption{\textsc{Definitions of Key Symbols Used in this Paper.}}
\label{tab:Table1}
\resizebox{\columnwidth}{!}{%
\begin{tabular}{|l|l|}
\hline
Symbol      & \multicolumn{1}{c|}{Definition} \\ \hline
$L$           & The total number of RATs.                               \\ \hline
$U$           & The total number of EDs.                                \\ \hline
$L_u$         & The subset of RATs assigned to the $u$th ED.              \\ \hline
$U_l$         & The subset of EDs assigned to the $l$th RAT.              \\ \hline
$x_{lu}$        & The assignment indicator for $u$th ED over $l$th RAT.                  \\ \hline
$\mathbb{U}_{lu}$ & The utility function of $u$th ED over $l$th RAT                     \\ \hline
$R_{lu}$        & The data rate for $u$th ED over $l$th RAT.                  \\ \hline
$R_{u}$        & The achieved rate by $u$th ED from its assigned RATs.                  \\ \hline
$R_{u}^{min}$ & The minimum required rate for $u$th ED.                   \\ \hline
$C_{lu}$       & The monetary cost per second for $u$th ED over $l$th RAT.              \\ \hline
$g_{lu}$        & The channel gain for $u$th ED over $l$th RAT.               \\ \hline
$\Gamma _{lu}$   & The SNR for $u$th ED over $l$th RAT.                         \\ \hline
$p_{lu}$        & The allocated power for $u$th ED over $l$th RAT.            \\ \hline
$P_{l}^{max}$/$W_l$ & The total power/bandwidth of $l$th RAT.                   \\ \hline
$h_{lu}$        & The small scale fading for $u$th ED over $l$th RAT.         \\ \hline
$\varepsilon_l$         & The monetary cost per bit for $l$th RAT.                  \\ \hline
$\alpha_u$ \& $\gamma_u$ & \begin{tabular}[c]{@{}l@{}}The weighting coefficients representing relative\\ importance of the two metrics of utility function.\end{tabular} \\ \hline
\begin{tabular}[c]{@{}l@{}}$\zeta_{ES},\eta_{ES},$ \\ $\eta_1, \eta_2, \zeta_l$ \end{tabular}& \begin{tabular}[c]{@{}l@{}}The weighting factors used in rewards to show if satisfying \\ the constraints has priority over maximizing objective.\end{tabular}  \\ \hline
\end{tabular}% 
}
\vspace{-4mm}%Put here to reduce too much white space after your table
\end{table}

Our objective is to study the problem of downlink sum-rate maximization of the multi-RAT HetNet while;
\begin{enumerate}
    \item reducing the monetary cost of connection,
    \item assigning each ED to the optimal set of RAT(s),
    \item allocating the optimal downlink power for each RAT-ED communication link,
    \item meeting the QoS requirements of EDs in terms of the minimum required data rate.
\end{enumerate}

We denote by $\cal L$ $\triangleq \{ 1, 2, \cdots, l, \cdots, L \}$ the set of all RATs indices, where $L$ represents the total number of RAT APs. We also denote by $\mathfrak{U}$ $\triangleq \{1, 2, \cdots, u, \cdots, U \}$, the set of all EDs indices, where $U$ represents the total number of multi-homing EDs. The subset of EDs assigned to the $l$th RAT is denoted by $U_l$ such that $U_l \subseteq$ $\mathfrak{U}$ $\forall l \in$ $\cal L$. In addition, the subset of RATs assigned to the $u$th ED is denoted by $L_u$ such that $L_u \subseteq$ $\cal L$ $\forall u \in$ $\mathfrak{U}$.

Assuming an OFDM-based system with flat fading for each ED and the total bandwidth of each RAT is equally divided between the RATs' assigned EDs, i.e., $W_l/U_l$, the upper bound of downlink data rate for the $u$th ED over the $l$th RAT at time slot $t$ is expressed as \cite{alwarafy2021DeepRAT, ciftler2021dqn, RN282}:
\begin{eqnarray}
R_{lu}(t) = \frac{W_{l}}{U_l} \mathrm{log_{2}} \Bigg( 1 + \frac{g_{lu}(t) p_{lu}(t)}{ \frac{W_{l}}{U_l} \sigma_{l}^{2}} \Bigg),
\label{rate_Eq}
\end{eqnarray}
where $W_{l}$ is the total bandwidth of the $l$th RAT and $\sigma_{l}^{2}$ is the $l$th RAT power spectral density of the additive white Gaussian noise (AWGN). The parameter $g_{lu}(t)$ in (\ref{rate_Eq}) represents the channel gain for the $u$th ED over the $l$th RAT at time slot $t$, which is defined based on the channel model of the RAT used. In Section \ref{sec4}, we will simulate three types of channel models, namely, the mmWave with beamforming for 5G NR \cite{Mismar2021Joint}, the COST 231 for 4G-LTE \cite{Mismar2021Joint}, and the exponential for 3G \cite{RN282, RN297}. Also, we will simulate a dynamic wireless system, as we will discuss later in Section \ref{sec4}.

Now we define $\varepsilon_l$ as the monetary cost of the $l$th RAT, which is expressed in Euro per bit \cite{chkirbene2021deep} and can be obtained from e.g., the IEEE 802.21 standard \cite{ieee2009ieee}. In our system model, such information can be easily gathered by the ES and stored in advance and only updated if there are changes in RATs' pricing. The monetary cost, expressed in Euro per second, resulting from using the $l$th RAT by the $u$th ED to receive the data rate $R_{lu}(t)$ at time slot $t$ is expressed as $C_{lu}(t) = \varepsilon_l R_{lu}(t)$. Table \ref{tab:Table1} summarizes the definitions of the symbols used in this paper. 

In this paper, we are optimizing the downlink sum-rate of the HetNet in a cost-effective way, such that the QoS requirements of EDs are guaranteed. This is achieved via jointly : 1) assigning the optimal set of RAT(s) to each ED, i.e., $x_{lu} \forall u \in \mathfrak{U}$, $l \in$ $\mathcal{L}$, and 2) allocating the optimal downlink power to each active RAT-ED link, i.e., $p_{lu} (t),~ \forall l \in$ $\mathcal{L}$, $u \in U_l$. Consequently, our optimization problem is formulated as follows: 
\begin{equation}
\begin{aligned}
\textbf{P1:}~~~~ \underset{x_{lu}, p_{lu}}{\max}
& ~~~ \sum_{l=1}^{L} {\sum_{u =1}^{U} {x_{lu}\mathbb{U}_{lu}(t)}} \\
\text{subject to}   \quad
& C_1: ~~ \sum_{l =1}^{L} x_{lu} \geq 1, ~ x_{lu} \in \{0, 1\}   \quad \forall u \in \mathfrak{U}, \\
& C_2: ~~ \sum_{u = 1}^{U} {x_{lu} p_{lu}(t)} \leq P_{l} ^ {max} \quad \forall l \in \mathcal{L}, \\
& C_3: ~~ \sum_{l =1}^{L} x_{lu} R_{lu}(t) \geq R_{u} ^ {min} \quad \forall u \in \mathfrak{U}, \\
& C_4: ~~ p_{lu}(t) \geq 0,~ \quad \forall l \in \mathcal{L},u \in \mathfrak{U}.
\label{Eq5}
\end{aligned}
\end{equation}
where $x_{lu}$ is the RTAs-EDs assignment indicator, such that $x_{lu}=1$ if the ES assigns the $l$th RAT to the $u$th ED, and $x_{lu}=0$ otherwise. To achieve our goal, we combine the data rate and monetary cost as a weighted sum objective utility function. In the weighted sum method, the Pareto optimal values can be achieved by adjusting the weighting parameters \cite{zhou2011multiobjective}, and thus there is no optimality loss in the problem formulation. Therefore, we define $\mathbb{U}_{lu}(t)$ in \textbf{P1} as the utility function of the $u$th ED over the $l$th RAT at time slot $t$, which is given by:
\begin{eqnarray}
\mathbb{U}_{lu}(t) = \alpha_{u} R_{lu}(t) - \gamma_{u} C_{lu}(t)
\label{Eq3_4}
\end{eqnarray}
where $\alpha_u$ and $\gamma_{u}$ are weighting parameters representing the relative importance of the objectives of jointly maximizing data rate and reducing the cost at each ED, such that $\alpha_{u} + \gamma_{u} = 1$. In other words, $\alpha_u$ and $\gamma_u$ are EDs'-defined parameters used to show if the ED cares more about getting a higher data rate over the monetary cost or getting a lower monetary cost over the data rate. Our objective is to find the optimum data rates $R_{lu}(t)$ that maximize our utility function via jointly controlling the RATs-EDs assignment and links' power allocations while considering the EDs preferences in terms of $\alpha_u$, $\gamma_u$, and $R_u^{min}$ and the RATs' rate prices $\epsilon_l$. Note that all quantities in (\ref{Eq3_4}) are normalized to their maximum values in order to make them comparable.

The optimization problem (\ref{Eq5}) is over the two unknowns: $x_{lu}$ and $p_{lu}$ subject to four constraints. $C_1$ ensures that each ED can be connected to multiple RATs simultaneously, and it reflects the multi-homing capabilities of EDs. $C_2$ and $C_4$ ensure that the power allocations from RATs to their assigned EDs do not violate the RATs' available power resources. $C_3$ ensures that the achievable data rates for EDs from their assigned RATs are greater than the minimum QoS requirements. 

\subsection{Why DRL? } \label{why_DRL}
Problem (\ref{Eq5}) is a combinatorial Mixed-Integer Non-Linear Programming (MINLP) \cite{boyd2004convex}, which is highly complex to solve using traditional approaches. In particular, applying the exhaustive search algorithm to find $x_{lu}$ followed by optimization approaches to find the corresponding $p_{lu}$ is not practical as the search space will grow exponentially. For example, as we will show later in Section \ref{sec4}, we simulate a scenario with $L=3$ and $U=10$. This means that applying the exhaustive method requires a full search over $(2^L -1)^U = 282,475,249$ possible combinations, each of which is followed by a constrained optimization process to find $p_{lu}$. This is quite difficult and impractical. In addition, transforming the problem into a geometric program is not possible due to constraint $C_1$ with the nonlinearity of the objective \cite{abdellatif2021onsra, boyd2004convex}. Note that compared to the problem formulation in \cite{RN19}, in \textbf{P1} we add the monetary cost aspect to the utility function, the multi-homing capabilities of EDs, and the power allocation issue. These aspects added new dimensions to the formulated problem and increased its complexity. In addition, compared to the problem formulation in \cite{zhang2020intelligent}, we added the following four additional dimensions in \textbf{P1}: the monetary cost issue to the objective function, the multi-homing constraint, the RATs' power allocation constraint, and the QoS requirements constraint of EDs. These new dimensions enriched the problem while making it more computationally expensive and difficult to solve using conventional approaches (e.g., optimization, ranking-based, and game theory methods discussed in Section \ref{Introduction}). The same observations are made when comparing \textbf{P1} with the problems formulated in \cite{li2017user} and \cite{Chen2016Joint}. Furthermore, unlike the previous works in \cite{RN284, RN287}, we added the monetary cost issue and the multi-homing constraint in \textbf{P1}. Note that compared to our problem formulation in \cite{alwarafy2021DeepRAT}, we add the monetary cost and multi-homing dimensions. In addition, this paper carefully addresses the scalability issue of the proposed DeepRAT algorithm for the increasing number of EDs. Finally, we demonstrate how our proposed DRL-based algorithm can quickly adapt, in terms of convergence speed, to the abrupt changes of the network, such as EDs' mobility.

Hence, and due to the high complexity of our formulated optimization problem, we propose to solve it using emerging DRL techniques instead. In particular, we hierarchically decompose \textbf{P1} into two optimization sub-problems, such that each sub-problem is a function of only one decision variable and, hence, can be solved separately and independently of the other sub-problem. The first sub-problem is to find the optimal RAT-EDs assignment $x_{lu}$, which depends on the parameters of the EDs. It is considered a global variable relevant to the overall HetNet system, and it can be solved at the ES level. The second sub-problem is to find the optimal power allocation for each RAT-ED $p_{lu}(t)$. It is considered a local variable that depends only on the parameters of RATs and can be solved at the RATs level. Consequently, \textbf{P1} is decomposed into the following two optimization sub-problems:
\begin{equation}
\begin{aligned}
\textbf{SP1:}~~~~ \underset{x_{lu}}{\max}
& ~~~ \sum_{l=1}^{L} {\sum_{u =1}^{U} {x_{lu}\mathbb{U}_{lu}(t)}} \\
\text{subject to}  \quad C_1 \quad \text{and} \quad C_3.
\label{Eq5_SP1}
\end{aligned}
\end{equation}
and 
\begin{equation}
\begin{aligned}
\textbf{SP2:}~~~~ \underset{p_{lu}}{\max}
& ~~~ \sum_{l=1}^{L} {\sum_{u =1}^{U} {x_{lu}\mathbb{U}_{lu}(t)}} \\
\text{subject to}  \quad  C_2 - C_4.
\label{Eq5_SP2}
\end{aligned}
\end{equation}

Next, we proceed with our proposed methodology to solve sub-problems \textbf{SP1} and \textbf{SP2} using DRL.

\section{DRL for Dynamic Multi-RAT assignment and Power Allocation} \label{DeepRAT_Architcutre}
In this section, we first explain our proposed DeepRAT algorithm to solve sub-problems \textbf{SP1} and \textbf{SP2}. Then, we provide a detailed description of the elements of the DQN and DDPG models used in the proposed DeepRAT model.

\subsection{The DeepRAT Framework}
We propose a multi-agent DRL-based framework called the DeepRAT, which hierarchically solves \textbf{SP1} and \textbf{SP2} iteratively and interactively in two stages; \emph{RATs-EDs assignment} and continuous \emph{power allocation}. DeepRAT employs two types of DRL algorithms, a single-agent DQN at the ES and multi-agent DDPG at the RATs level, as depicted in Fig. \ref{Fig1}. The methodology of the DeepRAT algorithm to solve the two sub-problems is explained as follows. For sub-problem \textbf{SP1}, DeepRAT utilizes a single-agent DQN algorithm to optimize the RATs-EDs assignment $x_{lu}$ at the ES level, while considering $p_{lu}$ as constants, which are passed by the RATs. Note that this RATs-EDs assignment is initially performed randomly by the ES without prior knowledge of whether it would be optimal or not. Then, the ES broadcasts $x_{lu}$ to the multi-agent DDPG algorithms of each RAT in order to optimize their power allocation $p_{lu}$ according to sub-problem \textbf{SP2} while considering $x_{lu}$ as constants. The ES then receives feedback ACK signals from all RATs indicating whether the objective of \textbf{P1} has been successfully solved for the current RATs-EDs assignment $x_{lu}(t)$ (i.e., the RATs are in good status) or not (i.e., the RATs are in bad status). Based on these ACK signals, the ES starts learning to make better assignments in the future time slots. These two stages are iteratively executed until all DeepRAT's agents learn the global policy that solves our main problem in \textbf{P1}, i.e., the single-agent DQN learns the optimal RATs-EDs assignment policy, and the multi-agents DDPG learn the optimal power allocation policy. The main elements of these two types of DRL models are defined next.

\subsection{DeepRAT Stage 1: DQN Algorithm for RATs-EDs Assignment} \label{DQN_archtitcute}
Due to the discrete nature of the RATs-EDs assignment problem, we adopt the DQN algorithm to act as an ES agent to learn the optimal policy for this problem. Below, we define the state space, action space, and reward function for the single-agent ES DQN model. 

\subsubsection{DQN action space:}
At each time step $t$, the main role of the DQN ES agent is to take an action $a^{ES}_t$ that optimally assigns each of the EDs to the optimal set of RAT(s), i.e., obtaining $x_{lu}$ in \textbf{SP1}. As shown in the optimization problem \textbf{SP1}, the DQN assignment action should: 1) maximize the objective function, and 2) satisfies $C_1$ and $C_3$. These conditions can be achieved iteratively by the design of the reward function. This action is a combinatorial problem that scales exponentially with the number of EDs and RATs, causing degradation in both system scalability and convergence speed. Unlike our previous work \cite{alwarafy2021DeepRAT}, in which the size of action space was proportional to both the number of EDs and RATs, i.e., $L^U$, in this paper we carefully address this scalability issue. Specifically, we make the size of the action space proportional to only the number of RATs, i.e., $2^L$. The scalability enhancement achieved is evident. As an example, for $L=3$ and $U=5$, the size of the action space proposed in this paper is around $30$ times less than the one proposed in \cite{alwarafy2021DeepRAT}. Similarly, when $L=4$ and $U=10$, the size reduction of the action space is around $65536$. This will greatly enhance system scalability and reduce convergence time. Therefore, the action space of the DQN ES agent is discrete, corresponding to assigning the optimal set of RAT(s) $L_u$ to the $u$th ED $\forall u \in$ $\mathfrak{U}$, where $L_u \in$ $\cal L$, which is expressed as:
\begin{eqnarray}
& {\cal A}_{ES} (t) = [a_{1}(t), a_{2}(t), \cdots, a_{u}(t), \cdots, a_{U}(t)], \\
& a_{u}(t)  = \{L_1, L_2, \cdots, L_{2^L} \}, \nonumber
\label{Eq6}
\end{eqnarray}

\subsubsection{DQN state space:} 
Due to the holistic view of the DQN ES agent, its state space must include effective and rich information about RATs and EDs to help the DQN agent in taking optimal assignment actions. To address the scalability issue mentioned previously, the RATs-EDs assignment is done by the ES iteratively for each ED, i.e., the ES assigns the $u$th ED to the best set of RATs $L_u$, while considering the assignment of the remaining EDs constant. Therefore, the ES DQN is configured to run on episodes of $U$ EDs time steps, and the DQN state must indicate the ED investigated \cite{chkirbene2021deep}. In addition, unlike our previous work in \cite{alwarafy2021DeepRAT} in which we assumed that the CSI is available at the agents, in this work we consider a more practical and challenging scenario by assuming that the agents have limited information about network dynamics and CSI. In particular, we assume that only historical information about the achieved data rates is available to the agents. With this in mind, the state space of ES is comprised of two main components, global information related to all RATs and EDs in the network and local information related to the $u$th ED investigated. The global information contains three elements; the matrix of all $x_{lu}$ at the previous time step (i.e., $x_{lu}(t-1) \in \mathbb C^{U \times L} ~ \forall l, u$), the matrix of all $R_{lu}$ at the previous time step (i.e., $R_{lu}(t-1) \in \mathbb C^{U \times L} ~ \forall l, u$), and the vector of all $R_{u}^{\text{min}} \in \mathbb C^{U \times 1} ~ \forall u$. The local information is only related to the $u$th ED under investigation, which has three \textit{scalar} elements; the index of ED under investigation at the current time step $uth(t) \in \mathbb C^{1 \times 1}$, the minimum required date of this ED $R_{uth}^{\text{min}} \in \mathbb C^{1 \times 1}$, and the achieved data rate of this ED from \textit{its assigned RATs} at the previous time step $R_{uth}(t-1) \in \mathbb C^{1 \times 1}$. Consequently, the state space of the DQN ES agent is expressed as:
%\begin{equation} 
%\begin{split}
\begin{eqnarray}
{\cal S}_{ES}(t) = [\text{Global information}, ~ \text{Local information}], \\ \nonumber
{\cal S}_{ES}(t) = [\underbrace{x_{lu}(t-1), R_{lu}(t-1), R_{u}^{min}}_\text{Global information}, \\ 
 \underbrace{uth(t), R_{uth}^{\text{min}}, R_{uth}(t-1)}_\text{Local information}]. \nonumber
\label{state_ES}
\end{eqnarray}
%\end{split}
%\end{equation}

\subsubsection{DQN reward function:}
The reward function is designed to incorporate the objective of our optimization problem in \textbf{P1} on one hand and the constraints $C_1$ and $C_3$ on the other hand. Hence, the agent will receive a negative punishment if the constraints are violated. The instantaneous reward $r_{ES}$ that the ES agent receives when taking action $a_{ES}$ given state $s_{ES}$ for the $u$th ED is expressed as:
\begin{eqnarray}
 r_{ES} (t) \triangleq \eta_{ES} constraints + \zeta_{ES} objective, \\ \nonumber
 r_{ES} (t) = \eta_{ES} \sum_{u}^{U} \Big( \sum_{l \in L_u} R_{lu}(t) - R_{u}^{min} \Big) + \\ 
\zeta_{ES} \sum_{l=1}^{L} {\sum_{u =1}^{U} {x_{lu}\mathbb{U}_{lu}(t)}}. \nonumber
\label{ES_reward}
\end{eqnarray}
where $\eta_{ES}$ and $\zeta_{ES} $ are weighting factors used to indicate whether satisfying the constraints has priority over maximizing the objective or not. These factors are manually tuned during simulation.

The RATs-EDs assignment problem can be formulated based on the immediate rewards achieved. Towards this goal, the expected accumulated discounted instantaneous reward over the time horizon $T$ is defined as ${\cal R}_{ES} = \mathbb{E} \left[ \sum_{t=1}^{T} {\gamma^ {t-1} r_{ES} (t)} \right], ~$ where $0\leq \gamma \leq1$ is a discounted factor \cite{sutton2018reinforcement}. The objective of the DQN ES agent is to obtain the optimal decision policy $\pi_{ES} ^{*}$ (i.e., selecting the optimal RATs-EDs assignment $x_{lu}$) that maximizes ${\cal R}_{ES}$. This is expressed as $\pi_{ES} ^{*} = \underset{\pi}{\text{argmax}} ~ {\cal R}_{ES}$.% =$ $\underset{\pi}{\text{argmax}} ~ \mathbb{E} \left[ \sum_{t=1}^{T} {\gamma^ {t-1} r_{ES} (t)} \right]$.

However, as we discussed earlier in Subsection \ref{why_DRL}, this RATs-EDs assignment problem is space-hard and it is quite difficult for traditional resource allocation techniques to solve it \cite{alwarafy2021deep,RN1}. Therefore, the DQN algorithm can be leveraged instead to learn $\pi_{ES} ^{*}$. In DQN, the optimal policy is expressed as $ \pi_{ES}^* = \arg\max \limits_{a} Q_{ES} ^{\pi_{ES}^*} (s^{ES}_t, a^{ES}_t)$, where the function $Q_{ES} ^{\pi_{ES}^*} (s^{ES}_t, a^{ES}_t)$ is called the state-action value function. This value function defines the expected accumulated discounted instantaneous reward achieved when executing action $a^{ES}_t$ in state $s^{ES}_t$ and then following the policy $\pi_{ES}$ thereafter. The value-function is defined as $Q_{ES}^{\pi_{ES}^*} (s^{ES}_t, a^{ES}_t) = \mathbb{E}_{\tau ~ \pi_{ES}} [ {\cal R}_{ES} | {s^{ES}_t,a^{ES}_t}]$, and the DQN algorithm utilizes the following iterative Bellman equation to compute it:
\begin{equation}
Q_{ES}^*(s^{ES}_{t},a^{ES}_{t}) = r_{ES} (s^{ES}_{t},a^{ES}_{t}) + \gamma \max \limits_{a^{ES}_{t}} Q_{ES} ^ {*} (s^{ES}_{t+1}, a^{ES}_{t+1}),
\label{Eq11}
\end{equation}

At each decision time step $t$, the deep neural network (DNN) of the ES DQN model iteratively updates its weights $\theta_{ES}$ to minimize the following loss function:
\begin{equation}
 L_{ES}(\theta_t) = \mathbb{E}_{{s, a, r, s'} \in \mathcal{D}_{ES}} [(y_{ES} (t) {-} Q_{ES}(s^{ES}_t, a^{ES}_t | \theta_{ES}))^2],
\label{los_ES}
\end{equation}
where $y_{ES} (t) = r_{ES} (s^{ES}_t, a^{ES}_t) + \gamma \max \limits_{a^{ES}_{t+1}} Q_{ES} \left(s^{ES}_{t+1}, a^{ES}_{t+1} | \theta_{ES}^{'} \right)$  is the target value, which is obtained from the target network with old weights $\theta^{'}_{ES}$, and $\mathcal D_{ES}$ represents the DQN replay buffer. 

%
%\begin{equation}
%y_{ES} (t) = r_{ES} (s_t, a_t) + \gamma \max \limits_{a_{t+1}} ⁡Q_{ES} \left( s_{t+1}, a_{t+1} | \theta_{ES}^{'} \right)
%\label{Eq13}
%\end{equation}

\subsection{DeepRAT Stage 2: DDPG Algorithm for Power Allocation} \label{sec3-2}
The DDPG is an efficient DRL algorithm developed to learn policies for continuous-based problems with high dimensionality in state and action spaces \cite{lillicrap2015continuous, RN296, RN295}. The DDPG algorithm will be leveraged in our second stage, i.e., solving the power optimization problem. In particular, a multi-agent deployment is considered in which each RAT employs a DDPG agent, whose main goal is solving its own objective function in \textbf{SP2} for its assigned EDs, $U_l$. The main elements of the multi-agent DDPG algorithm used are defined below.

\subsubsection{DDPG action space:}
Each DDPG RAT agent takes action independently and uncooperatively from the other agents. At time slot $t$, once the DQN ES agent executes the assignment action for the $u$th ED ($L_u$), the main goal of each DDPG agent is to optimize the power allocation $p_{lu}$ in \textbf{SP2} $\forall u \in \mathcal {U}$ and $\forall l \in \mathcal {L}$. As shown in the optimization problem \textbf{SP1}, the power allocation action of the $l$th DDPG RAT agent should: 1) maximize the objective function, and 2) satisfy $C_2$, $C_3$, and $C_4$. These conditions can be achieved by incorporating them into the reward function. Consequently, the action space of the $l$th  DDPG RAT agent is continuous with a size of $U$, corresponding to deciding the optimal power allocation for each of the RATs-EDs communication links (i.e., $p_{lu} (t)$). The action space of the $l$th DDPG RAT agent is defined as:
\begin{equation}
\begin{split}
{\cal A}_{l} (t) = [p_{l1}(t), p_{l2}(t), \cdots, p_{lu}(t), \cdots, p_{lU}(t)] \\
\text{where}~~~ p_{lu}(t)  \in [0, P_{l} ^ {max}]  ~~ Watt,
\end{split}
\label{EqDDPG_actSpace}
\end{equation}

The action exploration-exploitation problem in the DDPG algorithm is addressed via adding some Gaussian or Ornstein-Uhlenbeck (OU) noise \textbf{$n_t$} to the selected action $a_{t}^{l}$ \cite{lillicrap2015continuous}.

\subsubsection{DDPG state space:}
We assume that the DDPG RAT agents do not cooperate, and there is no direct communication between them. The agents, however, have direct communication with the DQN ES, which has a holistic view of all DDPG agents and can coordinate them. This means that each DDPG agent can acquire information about the other agents via the ES. The state space of the $l$th DDPG RAT agent is designed to contain useful information on the underlying HetNet. Four main types of representative information are incorporated in each $l$th agent state space. The first type of information is discrete and is directly related to the current RATs-EDs assignment action of the ES agent, which is a vector of the set of EDs assigned to the $l$th RAT at the current time step ($U_l(t) \in \mathbb C^{U \times 1}$). The second type of information is continuous, which is the vector of $R_{u}^{min} \in \mathbb C^{U \times 1}, ~ \forall u \in U_l$. The third information is the vector of $R_{lu}$ for the $l$th RAT at the previous time slot (i.e., $R_{lu}(t-1) \in \mathbb C^{U \times 1}, ~ \forall u \in U_l$). The fourth information is the vector of $R_u$ at the previous time slot (i.e., $R_u(t-1) \in \mathbb C^{U \times 1}, ~ \forall u \in U_l$). Note that while $R_{lu}(t-1)$ denotes the downlink rate for a \textit{single} link between the $u$th ED and $l$th RAT at the previous time slot, the notation $R_u(t-1)$ denotes the rate achieved by the $u$th ED from \textit{its assigned RATs} $L_u$ at the previous time slot, i.e., $ R_u(t-1) = \sum_{l \in L_{u}} {R_{lu}(t-1)}$. Consequently, the state space of the $l$th DDPG RAT agent is represented as:
\begin{equation}
\begin{split}
&{\cal S}_l (t) = [U_{l}(t), R_{u} ^ {min},R_{lu} (t-1), R_u(t-1)]. 
\label{EqDDPG_stateSpace}
\end{split}
\end{equation}

% malik this one is also for long equations
\begin{table*}[h!]
\centering
\begin{equation}
\begin{split}
&r_{l} (t) \triangleq constraints + \zeta_l objective,   \\ 
%&r_{l} (t) = \eta_1 (Pl_{max} - P_{l}) + \eta_2 (\sum_{u \in U_l} R_u - R_{u}^{min}) + \zeta_l objective, \\
&r_{l} (t) = \eta_1 \Big(P_l^{max} - \sum_{u \in U_l} p_{lu}(t)\Big) + \eta_2 \sum_{u \in U_l} \Big(\sum_{l \in L_u} R_{lu}(t) - R_{u}^{min} \Big) + \zeta_l \mathbb{U}_{lu}(t).
\end{split}
\label{EqDDPG_rewardFunction_span}    
\end{equation}
\hrule
\end{table*}

%\begin{align}
 %&r_{l} (t) \triangleq constraints + \zeta_l objective,  \\ 
 %&r_{l} (t) = \eta_1 (Pl_{max} - P_{l}) + \eta_2 (\sum_{u \in U_l} R_u - R_{u}^{min}) + \\ \nonumber
 %& \zeta_l objective, \\ \nonumber
 %&r_{l} (t) = \eta_1 (Pl_{max} - \sum_{u \in U_l} p_{lu}(t)) + \eta_2 \sum_{u \in U_l} (\sum_{l \in L_u} R_{lu}(t)) - %R_{u}^{min}) + \zeta_l \mathbb{U}_{lu}(t). \nonumber
%\label{EqDDPG_rewardFunction}    
%\end{align}

\subsubsection{DDPG reward function:}
The reward of the $l$th DDPG RAT agent is expressed as a continuous function that is governed by the RAT's achieved constrained objective function. It is quantified by including the optimization constraints $C_2$, $C_3$, and $C_4$ of (\ref{Eq5}) into the reward function so that the instantaneous reward reflects whether the constraints are satisfied or not \cite{RN295, chkirbene2021deep}. The reward function is given by (\ref{EqDDPG_rewardFunction_span}). In (\ref{EqDDPG_rewardFunction_span}), $\eta_1, \eta_2$, and $\zeta_{l} $ are also weighting factors used to indicate whether satisfying the constraints has priority over maximizing the objective or not. These factors are manually tuned during simulation.

In this second stage of our problem \textbf{SP2}, the main objective is to derive the optimal power allocation policy $\pi_{l} ^ {*}$ that maximizes the long-term reward of the $l$th agent ${\cal R}_{l}$, i.e., $\pi_{l} ^{*} = \underset{\pi}{\text{argmax}} ~ {\cal R}_{l} = $  $ \underset{\pi}{\text{argmax}} \sum_{t = 0}^{T} {\gamma^ {t-1} r_{l} (s^l_t, a^l_t)} $. Towards this goal, we implement the DDPG algorithm to derive this policy $\pi_{l} ^ {*}$. The DDPG algorithm integrates the DQN and actor-critic algorithms \cite{lillicrap2015continuous, RN295}, and will be utilized to perform the training of the RATs' DNNs. The DDPG has one parameterized actor function and one parameterized critic function represented by $ \mu (s^l_t | \theta_l^ \mu)$ and $Q_l(s^l_t,a^l_t | \theta_l ^Q)$, respectively, where $\theta_l ^ \mu$ and $\theta_l ^ Q$ denote the weights of the actor and critic networks, respectively. The parameterized actor function is used to derive the policy, and it is implemented with a DNN trained based on the iterative Bellman equation. On the other hand, the parameterized critic function is used to derive the value function $Q_l ^ \mu (s^l_t,a^l_t)$, defined as $Q_l ^ \mu (s^l_t,a^l_t) = \mathbb{E}_{\tau ~ \pi} [ {\cal R}_{l} | {s^l_t,a^l_t} ]$, and it is implemented using a DQN. The goal of each DDPG is to find the optimal policy $\pi_l ^ {*}$ that maximizes the long-term reward ${\cal R}_{l}$ using $\pi_{l} ^ {*} = \arg\max \limits_{a} Q_{l} ^ {\mu ^ {*}} (s^l_t,a^l_t)$. The value function $Q_{l} ^ {\mu ^ {*}} (s^l_t,a^l_t)$ is derived iteratively using the Bellman equation similar to (\ref{Eq11}), and the policy $\pi_l ^*$ is found via training the DNN of the $l$th DDPG RAT agent to minimize the Bellman loss function given by the following formula:
\begin{equation}
L_{l}(\theta_l^Q) =  \mathbb{E}_{{s, a, r, s'} \in \mathcal{D}_l} [(y_l(t) - Q_l(s^l_t, a^l_t | \theta_l^Q) )^2]
\label{loss_DDPG}
\end{equation}
where $\mathcal D_l$ denotes the $l$th DDPG agent's replay memory and $y_l(t)$ represents the target value, which is derived from the target network and obtained from the following equation:
\begin{equation}
y_l(t) = r_{l} (s^l_t, a^l_t) +\gamma \max \limits_{a_{t+1}} Q_l \left (s^l_{t+1}, \mu_l (s^l_{t+1} | \theta_{l}^{\pi_{l}^{'}}) | \theta_{l}^{Q^{'}} \right)
\label{yl_DDPG}
\end{equation}
where $\theta_l^{Q ^{'}}$ denotes the weights of target critic network, which has the same architecture as the main $Q$-network. These weights are mainly used to make the training more stable, and they are periodically updated based on the weights of the main $Q$-network $\theta_l^{Q}$. 

The actor network of the $l$th DDPG agent is trained via applying the chain rule to the expected return from the cumulative reward distribution $J$ with respect to $\theta_l^\mu$ \cite{lillicrap2015continuous}:
\begin{equation}
\nabla_{\theta_l^\mu} J = \mathbb{E} ~ \left[ \nabla_{\theta_l^\mu} Q_l (s^l_t, a^l_t | \theta_l^Q) |_{{s=s_t}, {a=\mu(s_t, {\theta_l^\mu})}} \right]
\label{gradient_DDPG}
\end{equation}

%\setlength{\textfloatsep}{} %\setlength{\textfloatsep}{\textfloatsepsave}
%\lipsum[12]
%\addtolength{\voffset}{.03in}
\begin{algorithm}[bt!]%[H]
%\algsetup{linenosize=\tiny}
\small
\begin{algorithmic}[1]
\renewcommand{\algorithmicrequire}{\textbf{Input:}}
\renewcommand{\algorithmicensure}{\textbf{Output:}}
\REQUIRE $L, U, R_u ^{min}, \alpha_u, \gamma_u$.
\ENSURE  Optimal RATs-EDs assignment $x_{lu}$ \& power allocation $p_{lu}$.
\STATE \textbf{Initialization}: Set $t=0$ and initialize $\mathcal{D}$$_{ES}$ of ES DQN agent and $\mathcal{D}$$_l$ of RATs' DDPG agents, $\forall l \in$ $\cal L$.
\STATE Randomly initialize weights of ES's DQN ($\theta_{ES}$ \& $\theta_{ES}^{'}$), and RATs' DDPG ($\theta_l^\mu$, $\theta_l^{Q}$, $\theta_l^{\mu ^{'}}$, \&  $\theta_l^{Q ^{'}}$).
\STATE Initialize states $s_0$ of ES and RATs with initial observations.
%\\ \textit{Malik beginning of LOOP Process}
\FOR {episode = 1 to $M$}
    \FOR {ED $= 1$ to $U$}
        \STATE Generate a random number $x$ from $[0, 1]$
        \IF {$x \geq \epsilon(t)$}
        \STATE Choose action $a_t^{ES}$ from ES's action space ${\cal A}_{ES} (t)$ according to $\max \limits_{a \in {\cal A}_{ES}} Q_{ES} \left(s_{t}^{ES}, a_{t}^{ES} | \theta_{ES} \right)$.
        \ELSE
        \STATE Choose a random action $a_t^{ES}$ from ${\cal A}_{ES} (t)$.
        \ENDIF
        \STATE Observe state for ES $s_t^{ES}$ and perform action $a_t^{ES}$.
        \FOR {$t_{RAT} = 0$ to $K$}
            \STATE Observe state for each RAT $s_{t}^ l$ using (\ref{EqDDPG_stateSpace}) and take action $a_{t}^l$ from ${\cal A}_{l} ^ {p}$ with OU noise, i.e., $a_{t}^l = \mu_l (s_{t}^l | \theta_{l} ^{\mu_l}) + n_{t} ^l$.
            \STATE Receive reward $r_{l} (s_{t}^{l},a_{t}^{l})$ using (\ref{EqDDPG_rewardFunction_span}), observe $s_{t+1}^l$ using (\ref{EqDDPG_stateSpace}), and store transitions $(s_t^l, a_t^l, r_l, s_{t+1}^l)$ in $\mathcal{D}_l$.
            \STATE Sample random mini-batch of $N$ transitions $(s_i^l, a_i^l, r^l_i, s_{i+1}^l)$ from $\mathcal{D}_l$.
            \STATE Set $y_l (i)$ based on (\ref{yl_DDPG}).
            \STATE Update $\theta_l^Q$ by minimizing loss (from (\ref{loss_DDPG})): $ L_{l}(\theta_l^Q) =  \frac{1}{N} \sum_{i=1}^{N}{(y_l(i) - Q_l(s_i^l, a_i^l | \theta_l^Q) )^2}$.
            \STATE Update $\theta_l^\mu$ using sampled policy gradient (from (\ref{gradient_DDPG})): $\nabla_{\theta_l^\mu} J = \frac{1}{N} \sum_{i=1}^{N}{\nabla_{\theta_l^\mu} Q_l (s_i^l, a_i^l | \theta_l^Q) |_{{s=s_i}, {a=\mu(s_i, {\theta_l^\mu})}} } $
            \STATE Update weights of all RATs' DDPG target networks:
            \STATE $\theta_l^{Q^{'}} \leftarrow \tau \theta_l^{Q} + (1-\tau) \theta_l^{Q^{'}}$,
            \STATE $\theta_l^{\mu^{'}} \leftarrow \tau \theta_l^{\mu} + (1-\tau) \theta_l^{\mu^{'}}$
            \ENDFOR
        \STATE Receive reward $r_{ES}(s_t^{ES}$, $a_t^{ES})$ using (\ref{ES_reward}), observe $s_{t+1}^{ES}$ using (\ref{state_ES}), \& store $(s_t^{ES}, a_t^{ES}, r_{ES}, s_{t+1}^{ES})$ in $\mathcal{D}_{ES}$,
        \STATE Sample random mini-batch of $M$ transitions $(s_i^{ES}, a_i^{ES}, r_{ES, i}, s_{i+1}^{ES})$ from $\mathcal{D}_{ES}$.
        %\STATE Set $y_i^{ES}$ based on Eq. (\ref{Eq18}).
        %\STATE Set $ \hat{y}_i^{ES} = ⁡Q_{ES} \left( s_{i}^{ES}, a_{i}^{ES} | \theta_{ES}^{'} \right)$.
        \STATE Update weights of ES $\theta_{ES}$ to minimize loss in (\ref{los_ES}).
    \ENDFOR
\ENDFOR
\end{algorithmic}
\caption{The Multi-Homing DeepRAT Algorithm}
\label{Algorithm1}
\end{algorithm}
%\addtolength{\voffset}{.03in}    
%\setlength{\textfloatsep}{0pt}% Remove \textfloatsep

The detailed pseudo-code of our proposed multi-homing DeepRAT algorithm for solving (\ref{Eq5}) is given in Algorithm \ref{Algorithm1} and explained next. Lines 1 to 3 initialize the network parameters, ES DQN model, RATs DDPG models, and initial states. The episode begins with initial states for all the agents and iterates over all EDs. In Lines 4 to 12, the DQN ES agent observes the state for each ED and takes the corresponding assignment action to the RATs, $L_u$. In Lines 13 to 23, each DDPG RAT agent observes the state space (including the current assignment action of the DQN agent) and takes the corresponding power allocation action to the EDs $p_{lu} ~ \forall u \in U_l$. In Lines 24 to 27, the DQN ES agent receives the reward for each ED assignment and learns to take a better assignment in future episodes. This process is repeated until the DeepRAT converges to the optimal policy that solves our main problem \textbf{P1} in (\ref{Eq5}), i.e., the DQN ES agent learns the optimal RATs-EDs assignment policy and all the DDPG RAT agents learn the optimal power allocation policy. 

\subsection{Deployment Scenario of the DeepRAT Framework}
Our proposed multi-homing DeepRAT algorithm is simple yet practical for implementation using simple software-defined radios (SDRs). During the training phase of the DeepRAT model, the expensive computations are conducted offline on quasi-centralized hardware, such as GPUs and/or tensor processing units. Once the DeepRAT algorithm learns the optimal global policy, it can be deployed online to perform optimal decisions autonomously by relying only on its learned policies without inducing any extra delay. This aspect will be quantified in the next section. It is noteworthy that updating DeepRAT’s DNNs is only required if the characteristics of the wireless environment have changed significantly, and they are no longer reflecting the training experiences. Such a case occurs once per several weeks or even months.

%\vspace{-.5cm}
\begin{figure*}[h!] % rate of each ED
	\centering
	%\vspace{-.5cm}
	\includegraphics[width=18.3cm, height=5.5cm,center]{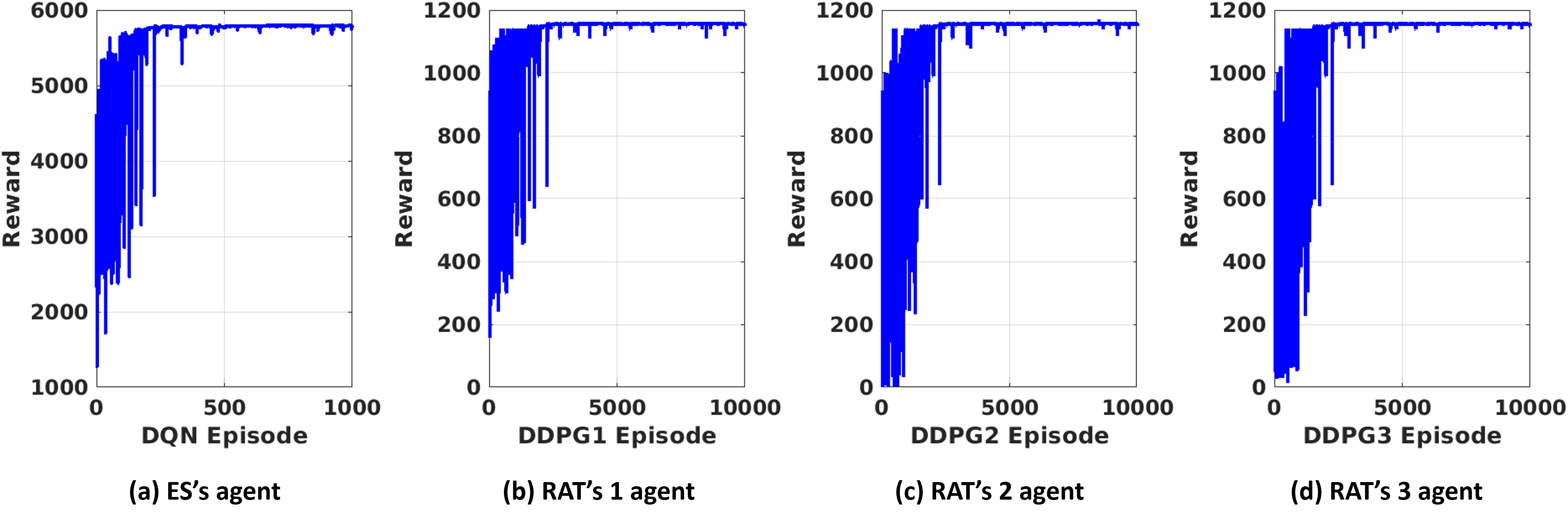}
	\caption{Reward training convergence of the DeepRAT's DQN and multi-agents DDPG models.}
	\label{Fig_allRewards}
\vspace{-5mm}%Put here to reduce too much white space after your table	
\end{figure*}

In addition, the design principle of the DeepRAT framework is quite practical for the modern AI-driven wireless networks. Here, we provide three practical deployment scenarios of the proposed DeepRAT ecosystem. First, the DeepRAT can be deployed in the Cloud/Centralized Radio Access Networks (C-RANs) architecture, in which the cloud-based ES can be allocated at the baseband unit (BBU) pool of the C-RANs. Second, the DeepRAT framework can be deployed in the Software-Defined Networking (SDN) architecture, in which the ES can be placed at the control-plane side of the architecture. Third, the DeepRAT framework can be deployed in the future self-organizing/sustaining networks \cite{RN234}, in which the ES can be placed at the self-organizing/sustaining server.

%\vspace{-.924cm}
\section{Performance Evaluation} \label{sec4}
This section presents a detailed description of the simulation setup we used to evaluate the performance of our proposed multi-homing DeepRAT algorithm. We first discuss the specifications of the HetNet under investigation, the DRL models used, and the EDs requirements. Then, we present and discuss the simulation results. Also, we present a practical scenario where there are abrupt changes in EDs mobility to demonstrate the ability of our proposed DeepRAT model to dynamically adapt to these varying system dynamics.

%\vspace{-.24cm}
\subsection{Simulation Setup} \label{sec4-a}
We consider a practical scenario of a next-generation cellular network comprised of three multi-RAT OFDM-based systems, i.e., $L=3$, specifically, 5G NR, 4G LTE, and 3G, as shown in Fig. \ref{Fig1}. The specifications of these three systems are shown in Table \ref{Table_RATs}. Note that during our simulation, we considered three practical channel models for each RAT with flat fading, namely, the mmWave with beamforming for 5G NR \cite{Mismar2021Joint}, the COST 231 for 4G-LTE \cite{Mismar2021Joint}, and the exponential for 3G \cite{RN282, RN297}. The specifications of these channel models are listed in Table \ref{Table_RATs}. The RATs are 100 meters apart. Ten single-antenna EDs are assumed, i.e., $U=10$, requesting services from the ES with random QoS requirements i.e., $R_{u} ^ {min}$, $\alpha_u$, and $\gamma_u$ as shown in Table \ref{Table_Ru_min}. In order to model a dynamic wireless system, we assume a time-varying network where the mobility of EDs varies over time with random speeds ranging from 2 to 6 km/h. This means that the CSI, in terms of channel gain, of all links will dynamically change over time.

The number of agents is four; one DQN-based located at the ES side and three DDPG-based located at each RAT, as depicted in Fig. \ref{Fig1}. These DRL models are simulated in Python using the Pytorch library, with architectures as shown in Table \ref{Table_Models_architicture}. Relu activation functions are used at the output layers of all NNs, and the weights are updated using the Adam optimizer \cite{kingma2014adam}. Also, in order to satisfy $C_4$ in \textbf{P1}, we employ the sigmoid function at the output layers of the DDPG actor networks. 

\begin{table}[t!]
\centering
\caption{\textsc{Simulation Parameters \cite{Mismar2021Joint, RN297}.}}
\label{Table_RATs}
\resizebox{\columnwidth}{!}{%
\begin{tabular}{|l|c|c|c|}
\hline
\textbf{Parameter} & \textbf{RAT1 (5G)} & \textbf{RAT2 (4G LTE)} & \textbf{RAT3 (3G)} \\ \hline
\textbf{Frequency (GHz)} & 28 & 6 & 2.4 \\ \hline
\textbf{Bandwidth (MHz)} & 200 & 40 & 27 \\ \hline
\textbf{Max power (dBm)} & 43 & 40 & 42 \\ \hline
\textbf{Noise spectral density (dBm/MHz)} & -57 & -57 & -57 \\ \hline
\textbf{Channel model} & Directional & COST 231 (Urban) & Exponential \\ \hline
\textbf{Path loss exponent} & 2(LOS), 4(NLOS) & - & 2 (LOS) \\ \hline
\textbf{Number of uniform linear array antennas} & 4 & 4 & 1 \\ \hline
\textbf{Number of multipaths} & 4 & 4 & - \\ \hline
\textbf{Antenna gain (dBi)} & 3 & 11 & - \\ \hline
\textbf{Shadowing (dB)} & 3.1 & 3 & 1.8 \\ \hline
\textbf{$\epsilon_{l}$ (Euro/bit)} & 9e-6 & 6e-6 & 1e-6 \\ \hline
\end{tabular}%
}
\vspace{-4mm}%Put here to reduce too much white space after your table
\end{table}

\begin{table}[tb!]
\centering
\caption{\textsc{QoS Requirements of EDs.}}
\label{Table_Ru_min}
%\resizebox{\columnwidth}{!}{%
\begin{tabular}{|l|l|l|l|}
\hline
\textbf{ED ID} & \textbf{$R_{u}^{min}$} (bps)& \textbf{$\alpha_u$} & \textbf{$\gamma_{u}$} \\ \hline
ED1 &  8.3 $\times 10^4$&  0.4& 0.6 \\ \hline
ED2 &  8.49 $\times 10^4$&  0.3& 0.7 \\ \hline
ED3 &  1.17 $\times 10^4$&  0.2& 0.8 \\ \hline
ED4 &  4.78 $\times 10^4$&  0.2& 0.8  \\ \hline
ED5 &  1.37 $\times 10^4$&  0& 1 \\ \hline
ED6 &  1.43 $\times 10^4$&  0.5& 0.5 \\ \hline
ED7 &  6.1 $\times 10^4$&  0.4& 0.6 \\ \hline
ED8 &  1.58 $\times 10^4$&  0.6& 0.4 \\ \hline
ED9 &  8.93 $\times 10^4$&  0.6& 0.4  \\ \hline
ED10 &  7.24 $\times 10^4$&  0.1& 0.9 \\ \hline
\end{tabular}%
%}
\vspace{-5mm}%Put here to reduce too much white space after your table
\end{table}

%\begin {center}
\begin{table}[tb!]
\centering
\vspace{.3in}
\caption{\textsc{Hyperparameters of the DeepRAT Model.}}
\label{Table_Models_architicture}
\begin{tabular}{|l|c|c|}
\hline
\multicolumn{1}{|l|}{\multirow{2}{*}{\textbf{Hyperparameter}}} & \multicolumn{2}{c|}{\textbf{Value}} \\ \cline{2-3}
\multicolumn{1}{|l|}{} & \multicolumn{1}{c|}{\textbf{Single-Agent DQN}} & \multicolumn{1}{c|}{\textbf{Multi-Agent DDPG}} \\ \hline
Buffer size & $1000$ & $500$ \\ \hline
Batch size & $64$ & $16$ \\ \hline
$\gamma$ & $0.99$ & $0.99$ \\ \hline
Number of layers & 2 & \begin{tabular}[c]{@{}c@{}}actor=$2$ \\ critic=$2$\end{tabular} \\ \hline
Number of neurons & (256, 128)& (16, 16) \\ \hline
Learning rate & $8 \times 10^{-4}$ & \begin{tabular}[c]{@{}c@{}}actor=$5 \times 10^{-4}$ \\ critic=$5 \times 10^{-4}$\end{tabular} \\ \hline
\begin{tabular}[c]{@{}l@{}}Exploration/\\Exploitation noise\end{tabular} &  \begin{tabular}[c]{@{}c@{}}$\epsilon_{start} = 1$\\ $\epsilon_{end} = 0.005$ \\ $\epsilon_{decay} = 5 \times 10^{-4}$\end{tabular} & \begin{tabular}[c]{@{}c@{}}OU ($\theta = 0.15$, \\ $\sigma = 0.03$)\end{tabular} \\ \hline
\begin{tabular}[c]{@{}l@{}}Rewards weighting \\ factors \end{tabular} & \begin{tabular}[c]{@{}l@{}}$\eta_{ES} = 1\times10^3$ \& \\ $\zeta_{ES} = 8\times 10^{-4}$ \end{tabular}& \begin{tabular}[c]{@{}l@{}} $\eta_1 = 1$, $\eta_2 = 1 \times 10^3$, \\ \& $\zeta_l = 5\times 10^{-3}$ \end{tabular}\\ \hline
\end{tabular}%
\vspace{-4mm}%Put here to reduce too much white space after your table
\end{table} 
%\end {center}

\begin{figure}[h!] % Utility or objective
	\centering
	%\vspace{-.5cm}
	\includegraphics[width=8.5cm, height=7.4cm,center]{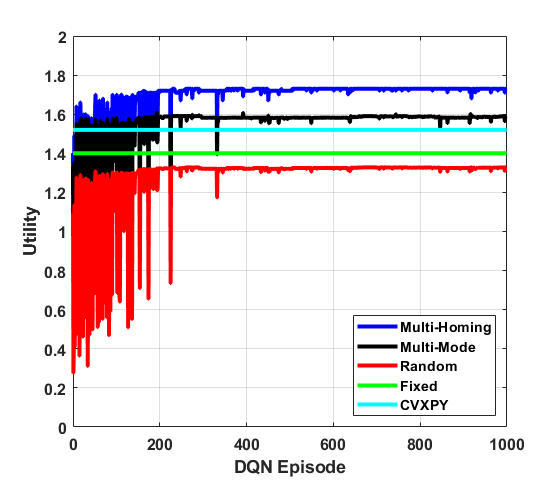}
	\caption{Utility of the proposed multi-homing DeepRAT algorithm compared to the multi-mode, random, fixed, and CVXPY schemes. Note that, unlike the proposed multi-homing DeepRAT technique, these four heuristic-based methods are not guaranteed to meet the EDs' QoS requirements in terms of rates, $\alpha_u$, and $\gamma_u$ preferences.}
	\label{Fig_utilitiy}
\vspace{-4mm}%Put here to reduce too much white space after your table	
\end{figure}

%\vspace{-2.cm}
\subsection{Numerical Results} \label{sec4-a}
In this subsection, we present simulation results to evaluate the performance of our proposed DeepRAT algorithm when deployed in an online fashion. Also, in order to evaluate the performance of our proposed multi-homing DeepRAT algorithm, we compare it against four state-of-the-art benchmarks. 1) The multi-mode method (i.e., maximum or greedy method), in which the ES will greedily assign each ED to only \textit{one} RAT that gives the maximum utility after the convergence of the DeepRAT algorithm, and the DDPG agents are utilized for power allocation, i.e., similar to our conference version in \cite{alwarafy2021DeepRAT} and the work in \cite{RN284}. It is noteworthy that the core difference between our implementation in this paper and the previous implementation in \cite{alwarafy2021DeepRAT} is the limited information about the multi-RAT network in this paper. In particular, our implementation of the multi-mode algorithm in \cite{alwarafy2021DeepRAT} assumes that the agents have full knowledge of network dynamics and CSI, i.e., channel gains $g_{lu}$, and we compared our approach against the CVXPY solver's solution \cite{diamond2016cvxpy}. However, our proposed multi-homing DeepRAT approach presented in this paper assumes that the agents have limited information about system dynamics and CSI). 2) The random approach, in which the ES will randomly assign each ED to only one RAT after the convergence of the DeepRAT algorithm, and the DDPG agents are used for power allocation. 3) The fixed approach, in which the ES will assign EDs to \textit{all} existing RATs, and the RATs will allocate their power \textit{equally} to all EDs. 4) The CVXPY solver's solution with MOSEK sub-solver \cite{diamond2016cvxpy}, in which the ES will assign EDs to \textit{all} existing RATs, and the CVXPY solver is utilized to solve the power allocation optimization problem assuming full knowledge of system dynamics and instantaneous CSI, similar to the works in \cite{RN296, alwarafy2021DeepRAT}. However, we should emphasize the following. 1) Although these conventional approaches show good results compared to our proposed multi-homing approach, they do not always guarantee an optimal solution, i.e., the QoS requirements of EDs are not guaranteed. 2) Unlike our proposed multi-homing DeepRAT approach, which works based on limited information about system dynamics and CSI, the fixed and CVXPY methods require perfect knowledge of the multi-RATs and instantaneous CSI. 3) Additionally, our DeepRAT algorithm adapts to abrupt network changes such as EDs' mobility. However, these conventional approaches are not adaptable, which severely degrades performance, accuracy, and reliability of the learned policies \cite{alwarafy2021deep, RN1}.

Fig. \ref{Fig_allRewards} shows the training rewards of all DeepRAT's DQN and DDPG agents. The DQN converges to the steady-state of optimal RATs-EDs assignment policy after 226 DQN episodes, while all three DDPG agents converge to the optimal power allocation policy after 2252 DDPG episodes. These results clearly show how our various value-based (i.e., DQN) and policy-based (i.e., DDPG) DRL agents efficiently interact with each other in order to learn a unified global optimal policy that solves our optimization problem in \textbf{P1}. 

\begin{figure}[t!] % CDF of utility
	\centering
	%\vspace{-.5cm}
	\includegraphics[width=8.7cm, height=7.2cm,center]{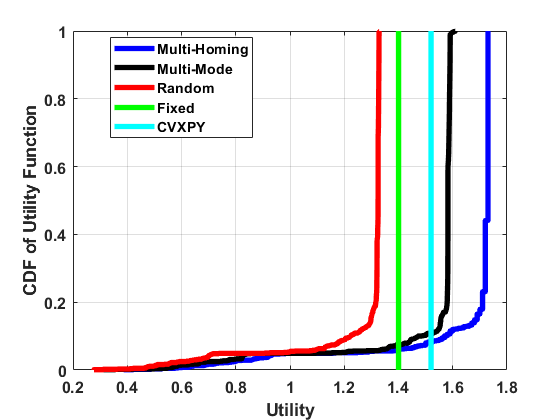}
	\caption{CDF of the utility function of the proposed multi-homing DeepRAT algorithm compared with the other heuristic-based schemes.}
	\label{Fig_CDF_of_Utility}
\vspace{-4mm}%Put here to reduce too much white space after your table
\end{figure}

Fig. \ref{Fig_utilitiy} shows the utility function, i.e., the objective of the optimization problem \textbf{P1}, which clearly demonstrates that the proposed multi-homing DeepRAT algorithm converges to the optimum solution after 226 episodes. Also, Fig. \ref{Fig_utilitiy} shows that the utility of the proposed multi-homing DeepRAT algorithm outperforms the ones achieved by the state-of-the-art approaches mentioned above, i.e., the multi-mode, random, fixed, and CVXPY approaches. In particular, the steady-state utility function of the proposed DeepRAT algorithm is 1.73 compared to 1.58, 1.52, 1.33, and 1.4, respectively, for the multi-mode, CVXPY, random, and fixed methods. Recall that the multi-mode, random, and fixed methods do not guarantee that they satisfy the EDs' QoS requirements. In addition, note that the multi-mode outperforms the CVXPY as the latter does not assign EDs to the optimal set of RATs leading to a lower utility value. Fig. \ref{Fig_CDF_of_Utility} also shows the corresponding cumulative distribution function (CDF) of the utility function for the proposed multi-homing DeepRAT algorithm and these four state-of-the-art approaches. It clearly shows that the median of the utility for multi-homing DeepRAT is 1.73 compared to 1.58, 1.52, 1.33, and 1.4, respectively, for the multi-mode, CVXPY, random, and fixed methods.

\begin{figure}[h!] % Utility or objective
	\centering
	%\vspace{-.5cm}
	\includegraphics[width=8.6cm, height=7.8cm,center]{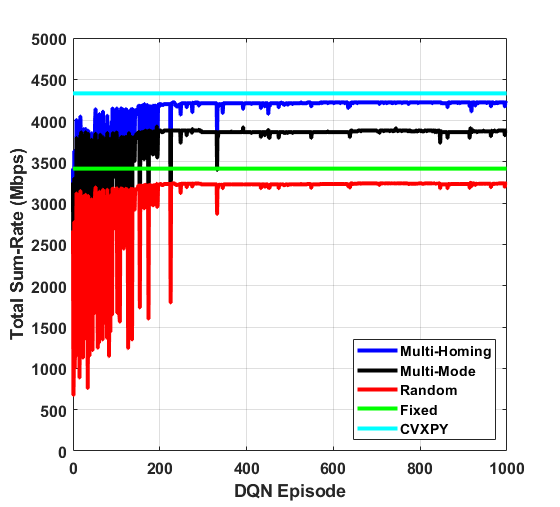}
	\caption{Total sum-rate achieved by the proposed multi-homing DeepRAT algorithm compared to the multi-mode, random, fixed, and CVXPY schemes. Although the CVXPY scheme provides a slightly better rate than DeepRAT, it does not assign EDS to the optimal set of RATs, requires full knowledge of CSI, and has inferior utility values as shown in Figs. \ref{Fig_utilitiy} and \ref{Fig_CDF_of_Utility}.}
	\label{Fig_R_Total}
%\vspace{-2mm}%Put here to reduce too much white space after your table
\end{figure}

In Fig. \ref{Fig_R_Total}, we present the total sum-rate achieved by our proposed multi-homing algorithm compared to the four conventional approaches. The steady-state of total sum-rate of our proposed DeepRAT approach is 4.2 Gbps compared to 4.33 Gbps, 3.86 Gbps, 3.23 Gbps, and 3.42 Gbps for the CVXPY, multi-mode, random, and fixed methods, respectively. Hence, the percentage increase in total sum-rate achieved by the multi-homing DeepRAT over the multi-mode, random, and fixed is 8.81\%, 30.03\%, and 22.81\%, respectively. Note that although the rate achieved by the CVXPY solver is slightly higher than the proposed DeepRAT algorithm, it has the following shortcomings. 1) Unlike the multi-homing DeepRAT method, the CVXPY-based approach does not guarantee that the EDs are assigned to the optimal set of RATs as it assigns all EDs to all existing RATs. 2) The CVXPY solver requires full and instantaneous knowledge of CSI to solve the power allocation problem, whereas the proposed DeepRAT method does not. This aspect is important when we discuss the fast adaptivity of our proposed DeepRAT algorithm in the next subsection. 3) The utility values obtained by the DeepRAT method are higher than those obtained by the CVXPY solver, as observed from Figs. \ref{Fig_utilitiy} and \ref{Fig_CDF_of_Utility}.

\begin{figure}[bt!] % rate percentage figure
	\centering
	%\vspace{-.5cm}
	\includegraphics[width=9cm, height=13cm,center]{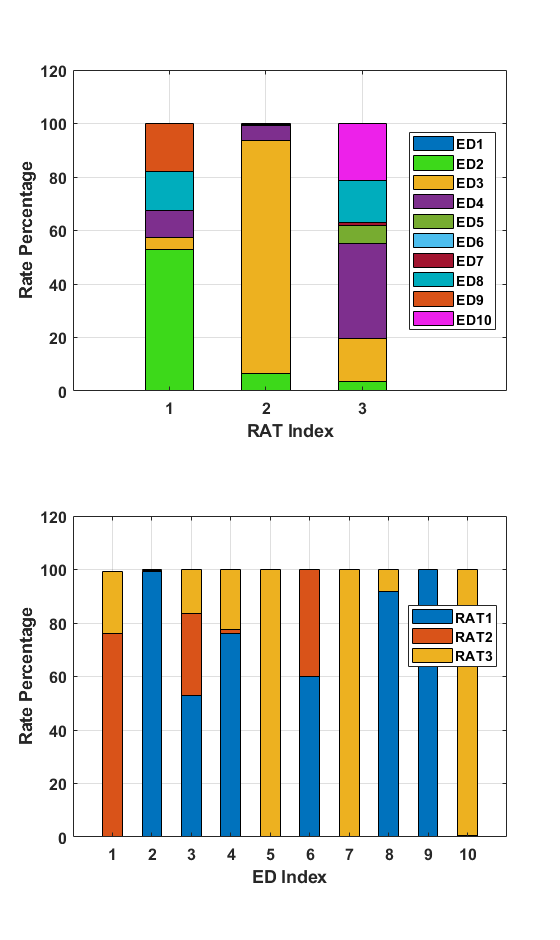}
	\caption{Percentage of data rate delivered from each RAT to its assigned EDs (top) and achieved by each ED from its assigned RATs (bottom).}
	\label{Fig_rate_percentage}
\vspace{-4mm}%Put here to reduce too much white space after your table
\end{figure}

The optimal RATs-EDs assignment process after the convergence of all  DeepRAT's models is shown in Fig. \ref{Fig_rate_percentage}. It shows which EDs have been assigned to RATs and the corresponding percentages of downlink data rates achieved. The top plot shows the percentage of downlink data rates delivered by each RAT to its assigned EDs, while the bottom plot shows the percentage of data rates delivered to each ED from its assigned set of RATs. As an example, the top plot shows that the ES assigned RAT1 to six EDs, namely ED2, ED3, ED4, ED6, ED8, and ED9 (i.e., $U_1 = \{2, 3, 4, 6, 8, 9\}$). The percentages of data rates delivered to these EDs from RAT1 are 52.9\%, 4.42\%, 10\%, 0.08\%, 14.7\%, and 17.9\%, respectively. Also, the bottom plot shows that the ES assigned ED3 to all three RATs (i.e., $L_3 = \{1, 2, 3\}$), and the percentages of data rates delivered from RATs 1, 2, 3 are 53\%, 30.7\%, and 16.3\%, respectively. For convenience, Table \ref{Table_percenetages} shows the results presented at the bottom of Fig. \ref{Fig_rate_percentage} in a tabular form.

\begin{table}[tb!]
\centering
\caption{\textsc{Percentages of Data Rates Delivered By RATs to Their Assigned EDs, as Shown at the Bottom of Fig. \ref{Fig_rate_percentage}.} } 
\label{Table_percenetages}
%\resizebox{\columnwidth}{!}{%
\begin{tabular}{|l|c|c|c|}
\hline
\textbf{ED ID} & RAT1 (\%)& RAT2 (\% )  & RAT3 (\% )  \\ \hline
ED1 & 0&76.2 & 23.1 \\ \hline
ED2 & 99.1 &0.36 & 0.54 \\ \hline
ED3 &  53& 30.7 &  16.3\\ \hline
ED4 & 76.2 & 1.3& 22.5 \\ \hline
ED5 & 0 & 0 &  100\\ \hline
ED6 & 60.1 & 39.9& 0 \\ \hline
ED7 & 0 & 0 & 100 \\ \hline
ED8 & 91.8 & 0 & 8.2 \\ \hline
ED9 & 100 & 0 &  0 \\ \hline
ED10 &  0& 0.7 &  99.3\\ \hline
\end{tabular}%
%}
\vspace{-5mm}%Put here to reduce too much white space after your table
\end{table}

\begin{figure}[t!] % rate of each ED
	\centering
	%\vspace{-.5cm}
	\includegraphics[width=9cm, height=6cm,center]{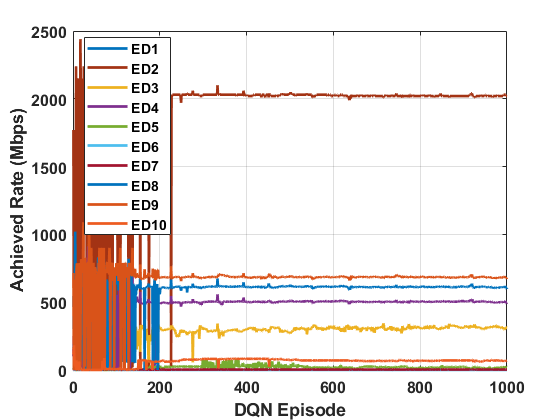}
	\caption{Total achieved rate by each multi-homing ED from its assigned RATs. All rates are greater than the minimum required data rate requested by EDs as shown in Table \ref{Table_Ru_min}.}
	\label{Fig_ratesOfEDs}
\end{figure}

Fig. \ref{Fig_ratesOfEDs} shows the achieved data rate for each multi-homing ED from its assigned RAT(s). All EDs converge to the optimal rate and reach steady-state after 226 episodes. Also, when comparing these results with the minimum data rate requirements of EDs $R_{u}^{min}$ in Table \ref{Table_Ru_min}, we can see that the ES assignment guarantees that all EDs satisfy their QoS data rate requirements in a manner that cost-effectively maximizes the network sum-rate. For example, the data rate requirement of ED4 is $47.8$ kbps, whereas the achievable rate after connecting ED6 to RATs 1 and 2 is 503 Mbps bps, which is much greater than the required rate. Indeed, our proposed scheme can significantly enhance the performance of the HetNet by enabling data transfer from RATs to EDs in a cost-effective manner while guaranteeing satisfactory QoS.

%\vspace{-.5cm}

%Malik Fig.2 total reward all episodes
%\begin{figure*}[bt!]
%    \begin{subfigure}{.24\textwidth}
%	\centering
%	%\vspace{-.5cm}
%	\includegraphics[width=4cm, height=4cm,center]{Figures2/reward_ES}
%	% \caption{Put your sub-caption here}
%	\label{reward_ES}
%	\end{subfigure}
%	\begin{subfigure}{.24\textwidth}
%	\centering
%    \includegraphics[width=4cm, height=4cm,center]{Figures2/reward_DDPG1}
%    % \caption{Put your sub-caption here}
%    \label{reward_DDPG1}
%    \end{subfigure}
%	\begin{subfigure}{.24\textwidth}
%	\centering
%    \includegraphics[width=4cm, height=4cm,center]{Figures2/reward_DDPG2}
%    % \caption{Put your sub-caption here}
%    \label{reward_DDPG2}
%    \end{subfigure}
%	\begin{subfigure}{.24\textwidth}
%	\centering
%    \includegraphics[width=4cm, height=4cm,center]{Figures2/reward_DDPG3}
%    % \caption{Put your sub-caption here}
%    \label{reward_DDPG3}
%   \end{subfigure}
%	\caption{Training convergence of the DeepRAT's DQN and multi-agents DDPG models.}
%	\label{Fig_training_rewards}
%\end{figure*}

\subsection{DeepRAT's Adaptivity to the Mobility of EDs} 
In this subsection, we demonstrate the efficiency of the DeepRAT algorithm to quickly and dynamically adapt, in terms of the convergence speed, to abrupt network changes. During the simulation, we define the convergence speed as the number of episodes required to reach the steady-state, which is defined as having the utility values constant for the last 200 episodes \cite{ciftler2021dqn}.

\begin{figure}[bt!] % rate of each ED 
	\centering
	%\vspace{-.5cm}
	\includegraphics[width=9cm, height=6cm,center]{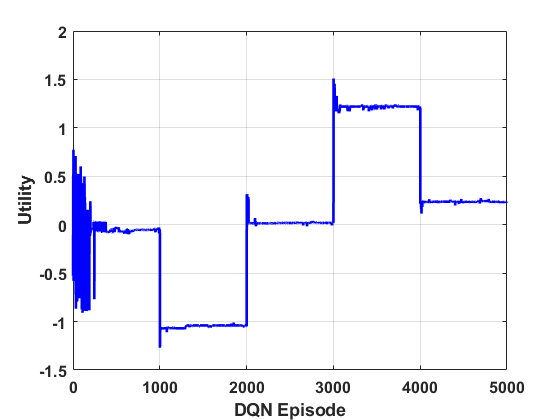}
	\caption{Adaptivity of the DeepRAT algorithm to the dynamic mobility of EDs. The EDs randomly move at every 1000 episodes, and the DeepRAT quickly adapts to these changes. The convergence speed after training is around 2.5 faster than the one at the initial training.}
	\label{Fig_adapt_mobility}
\vspace{-4mm}%Put here to reduce too much white space after your table
\end{figure}

%\textbf{Adaptivity Against EDs' Mobility:}\\
We investigate a practical scenario where the EDs move randomly during each 1000 episodes. Fig. \ref{Fig_adapt_mobility} shows the corresponding simulation results for the utility function. We notice that the DeepRAT algorithm adapts very quickly, in terms of the convergence speed, to the abrupt system dynamics, i.e., EDs mobility, and it dynamically finds the optimal solution of the problem in \textbf{P1}. These trends are clear at episodes 1000, 2000, 3000, and 4000, where the DeepRAT algorithm converges and reaches steady-states after 246, 1081, 2097, 3078, and 4034 episodes, respectively. The initial training phase takes around 246 episodes to converge, whereas the worst-case after the training takes only less than 97 episodes to converge, i.e., to re-solve the optimization problem \textbf{P1}. This means that the convergence speed after training is around 2.5 faster than the convergence speed at the initial training, which quantifies the dynamic adaption performance of the DeepRAT algorithm for the random changes in network dynamics.

% malik use this code to get the tikz-generated figure from python 
%\input{"E:/OneDrive - King Suad University/Research_Simulation/Books_Examples/tmp/stage2/Fig_TrainedReward1Episode_16_18_.tex"}

%\vspace{-0.16cm}
\section{Conclusion}\label{sec6} 
This paper investigated the problem of cost-effective downlink sum-rate maximization in multi-RAT multi-homing HetNets. The problem was formulated as a MINLP whose objective is to cost-effectively maximize network sum-rate via jointly assigning EDs to the optimal set of RATs and allocating the optimal RATs' power levels. Due to the high complexity and combinatorial nature of the problem on the one hand and the limited knowledge of network statistics on the other hand, we proposed to solve the problem using DRL methods. Towards this goal, we proposed a multi-agent DQN and DDPG-based DRL algorithm, called DeepRAT, which solved the problem hierarchically in two stages; dynamic RATs-EDs assignment and power allocation. Our simulation results showed that the proposed multi-homing DeepRAT algorithm outperforms four benchmark heuristic algorithms in terms of utility value. In addition, our simulation results showed the ability of the DeepRAT algorithm to quickly adapt to abrupt network changes, such as EDs' mobility, and that its convergence speed after training is around 2.5 faster than the initial training. As future work, we will extend the multi-homing DeepRAT algorithm to address the problem of joint optimization of both power and spectrum.

%\vspace{-.17cm}
\section*{Acknowledgment} 
This publication was made possible by NPRP-Standard (NPRP-S) Thirteen ($13^\text{th}$) Cycle grant \# NPRP13S-0201-200219 from the Qatar National Research Fund (a member of Qatar Foundation). The findings herein reflect the work, and are solely the responsibility, of the authors.
%%\nocite{*}
%%\vspace{-.1cm}
\vspace{-0.1cm}
\bibliographystyle{IEEEtran}
\bibliography{The_bibliography}

\end{document}